\documentclass[draft,prd,aps,showpacs]{revtex4}
\usepackage{amsmath,amssymb}
\usepackage{epsf}
\newcommand{\be}{\begin{equation}}
\newcommand{\ee}{\end{equation}}
\newcommand{\beq}{\begin{equation}}
\newcommand{\eeq}{\end{equation}}
\newcommand{\ba}{\begin{eqnarray}}
\newcommand{\ea}{\end{eqnarray}}

\newcommand{\nn}{\nonumber}

\newcommand{\bea}{\begin{eqnarray}}
\newcommand{\eea}{\end{eqnarray}}

\begin{document}
\title{\bf Non(anti)commutative superspace with coordinate-dependent deformation}
\author{L.G.~Aldrovandi,
F.A.~Schaposnik\footnote{F.A.S. is associated with CICBA.} and
G.A.~Silva\footnote{G.A.S. is associated with CONICET}}

\affiliation{Departamento\ de F\'{\i}sica, Universidad Nacional de
La Plata,
      C.C. 67, (1900) La Plata, Argentina.}
\date{\today}


\begin{abstract}
We  consider non(anti)commutative superspace with coordinate
dependent deformation parameters $C^{\alpha\beta}$. We show that a
chiral ${\cal N}=1/2$ supersymmetry can be defined and that chiral
and antichiral superfields are still closed under the Moyal-Weyl
associative  product implementing the deformation.  A consistent
${\cal N}=1/2$ Super Yang-Mills deformed theory can be constructed
provided $C^{\alpha\beta}$ satisfies a suitable condition which
can be connected with the graviphoton background at the origin of
the deformation. After adding matter we also discuss the Konishi
anomaly and the gluino condensation.
\end{abstract}
\maketitle

\section{Introduction}
Recently, a deformation of the supersymmetry algebra  attracted
much attention due to its connection with string dynamics in non
trivial RR backgrounds\cite{deBoer}-\cite{lerda2}, the
deformation parameter $C^{\alpha\beta}$ being related to a
constant graviphoton field strength.  Moreover,
 a relation between  $C$-deformed super Yang-Mills
theory \cite{Seiberg} and conventional ${\cal N}=1$ super
Yang-Mills theory (SUSY gluodynamics) has been signaled in
\cite{shifman}. According to this last work, such a
$C$-deformation turns to be related to a spectral degeneracy in
SUSY gluodynamics (which due to the planar equivalence can be
related to one flavor QCD).

It was conjectured in \cite{shifman} that, in fact, ${\cal N}=1/2$
supersymmetry should remain valid for a {coordinate-dependent}
 $C^{\alpha\beta}$ deformation. In this note, we investigate such a
possibility by analyzing a deformed algebra for the fermionic
coordinates $\theta$ with $C^{\alpha\beta}$ depending on the
chiral variable $y$. As it happens for constant $C^{\alpha\beta}$,
we shall see that the subalgebra satisfied by $Q_\alpha$ is
preserved when $C^{\alpha\beta} = C^{\alpha\beta}(y)$ so that a
chiral ${\cal N}=1/2$ supersymmetry can be defined. We also show
that chiral and antichiral superfields are still closed under the
Moyal-Weyl associative product implementing the deformation but
the case of antichiral superfields should be handled with care
  due to the fact that the chiral covariant derivative
$D_\alpha$ violates the Leibnitz rule. Moreover,   chiral and
antichiral  superfields strength do not in general transform
covariantly under general supergauge transformations. However, one
can still consistently define a super Yang-Mills deformed theory
by adopting, from the start, the Wess-Zumino gauge and
appropriately restricting   supergauge transformations. We show
that demanding gauge invariance of the resulting deformed theory
imposes a remarkable condition on the coordinate-dependent
$C^{\alpha\beta}$. Finally, we discuss the Konishi anomaly and the
gluino condensation for coordinate dependent deformation.

The paper is organized as follows. We introduce in section II the
coordinate dependent deformation, discuss how a Moyal-Weyl
associative product of superfields can be implemented and present
the supercharge algebra. Then, in section III we introduce chiral
and vector superfields carefully analyzing the condition under
which gauge invariant ${\cal N}=1/2$ supersymmetric Lagrangian can
be defined. Coupling the super Yang Mills multiplet to matter in
the fundamental, we present in section IV the supersymmetric
Lagrangian in component analyzing the essential features of the
deformed terms. We also discuss the Konishi anomaly in the
commutator leading to the gluino condensation showing that it
remains unchanged by the coordinate dependent deformation. Finally
we summarize and discuss our results in section V.

\section{Deformed superspace}

We consider the deformation of 4 dimensional Euclidean ${\cal N} =
1$ superspace parametrized by chiral bosonic coordinates $y^\mu
=x^\mu +i\hat \theta^\alpha \sigma^{\mu}_{\alpha
\dot\alpha}{\bar\theta^{\dot\alpha}}$ and chiral and antichiral
fermionic coordinates $\hat\theta^\alpha,{\bar\theta^{\dot
\alpha}}$ satisfying the Clifford algebra
\be
  \{\hat\theta^\alpha, \hat\theta ^\beta\} =
  C^{\alpha \beta}(y) \; , \;\;\; \;\;\;
  \{{\bar\theta^{\dot \alpha}},{\bar\theta^{\dot \beta}}\} = 0 \; , \;\;\; \;\;\;
  \{\hat\theta^\alpha,{\bar\theta^{\dot \beta}}\} = 0
  \label{uno}
\ee
where $C^{\alpha \beta}$ is some chiral coordinate-dependent
symmetric matrix. (We follow the conventions of ref. \cite{Wess}
for lowering and rising spinor indices.)
 We indicate with a hat that the
$\theta$ subalgebra is deformed. Following \cite{Seiberg} we also
define
\be
   [y^\mu,y^\nu] = [y^\mu,\hat \theta^\alpha] =
   [y^\mu,\bar\theta^{\dot\alpha}] = 0
\ee

Due to the non-anticommutativity of the $\hat\theta$'s
coordinates, functions in this deformed superspace have to be
ordered. One can show  that, as it is the case of constant
$C^{\alpha\beta}$, a Weyl ordering can be easily implemented \cite{zachos}.
Indeed, consider the Fourier transform $\tilde f$ of a function
$f$ in ordinary superspace given by
\be
  f(y ;\theta,  \bar\theta)=\int d^{2} \pi e^{- \pi  \theta}
  \tilde{f} (y ; \pi, \bar\theta )
\ee and define a one to one map between Weyl symbols $\hat f$ in
deformed superspace $(y ,\hat \theta,  \bar\theta)$ and functions
$f$  in ordinary superspace $(y , \theta,  \bar\theta)$ via the
formula
\be
  \hat{f}(y ;\hat \theta,  \bar\theta) \equiv \int d^{2} \pi e^{- \pi \hat \theta}
  \tilde{f} (y ;\pi,  \bar\theta )
\ee
The product $\cdot$ of symbols, $\hat {f}_1(y ,\hat \theta,
\bar\theta)\cdot\hat {f}_2(y ,\hat \theta,  \bar\theta)$, can then
be written as:
\bea
  \hat{f}_1( y ; \hat \theta, \bar \theta)\cdot \hat{f}_2(y ; \hat \theta,\bar\theta)
  &=& \int d^{2}\pi_1d^{2}\pi_2\,  e^{- \pi_1 \hat \theta}\cdot\,e^{- \pi_2 \hat \theta}
  \tilde{f}_1 (y; \pi_1,\bar \theta) \tilde{f}_2 (y;
  \pi_2,\bar\theta)\nonumber\\
  && =\int d^{2}\pi_1d^{2}\pi_2\, e^{- (\pi_1+\pi_2) \hat \theta}
  e^{-\frac 1 2 \pi_{1\alpha} C^{\alpha \beta}(y) \pi_{2\beta}}
   \tilde{f}_1 (y; \pi_1,\bar \theta) \tilde{f}_2 (y; \pi_2,\bar\theta)
   \label{prod}
\eea
where  the Baker-Campbell-Haussdorff formula has been used,
\be
   e^{- \pi_1 \hat \theta}\cdot\,e^{- \pi_2 \hat \theta}=
   e^{- (\pi_1+\pi_2) \hat \theta}
   e^{-\frac 1 2 \pi_{1\alpha} C^{\alpha \beta}(y) \pi_{2\beta}}
   \nonumber
\ee
After a change of integration variables, $\pi=\pi_1+\pi_2$,
$\pi'=\pi_1-\pi_2$   eq.(\ref{prod}) becomes
\bea
   \hat{f}_1\cdot \hat{f}_2(y; \hat \theta, \bar \theta)
   &=&\int d^{2}\pi d^{2}\pi'\, e^{- \pi \hat \theta}
   e^{-\frac 1 8 (\pi +\pi')_\alpha C^{\alpha \beta}(y)(\pi
   -\pi')_\beta}
   \tilde{f}_1 (y;\frac {\pi +\pi'}{2},\bar \theta) \tilde{f}_2
   (y;\frac {\pi -\pi'}{2},\bar \theta)
\eea
so that a product of symbols gives,
\be
   {\widetilde{{f}_1\cdot{f}_2}}(y;\pi,\bar \theta)= \int d^{2}\pi'\,
   e^{-\frac 1 8 (\pi +\pi')_\alpha C^{\alpha \beta}(y)(\pi
   -\pi')_\beta}\tilde{f}_1 (y;\frac {\pi +\pi'}{2},\bar \theta)
\tilde{f}_2
   (y;\frac {\pi -\pi'}{2},\bar \theta)
   \label{fourier1}
\ee
Using (\ref{prod}) it is easy to see that the product is
associative,
\bea
   \hat{f}_1\cdot(\hat{f}_2\cdot\hat{f}_3)
     & =&\int \!\!d^{2}\pi_1d^{2}\pi_2d^{2}\pi_3\,
    e^{- \pi_1 \hat \theta}\cdot\,e^{- (\pi_2+\pi_3) \hat \theta}
    e^{-\frac 1 2 \pi_{2} C \pi_{3}}   \tilde{f}_1(y;\pi_{1},\bar \theta)
    \tilde{f}_2(y;\pi_{2},\bar \theta)  \tilde{f}_3(y;\pi_{3},\bar \theta)\nonumber\\
    && \! \! \!\!\!\!\! \!  =\int d^{2}\pi_1d^{2}\pi_2d^{2}\pi_3\,
    e^{- (\pi_1+\pi_2+\pi_3) \hat \theta}
    e^{-\frac 1 2 (\pi_{1} C \pi_{2}+\pi_{1} C \pi_{3}+\pi_{2} C \pi_{3})}
   \tilde{f}_1(y;\pi_{1},\bar \theta)
    \tilde{f}_2(y;\pi_{2},\bar \theta)  \tilde{f}_3(y;\pi_{3},\bar \theta)\nonumber\\
    && \! \! \!\!\!\!\! \! =\int d^{2}\pi_1d^{2}\pi_2d^{2}\pi_3\,
    \,e^{- (\pi_1+\pi_2) \hat \theta}\cdot\,e^{- \pi_3 \hat \theta}
    e^{-\frac 1 2 \pi_{1} C \pi_{2}}\tilde{f}_1(y;\pi_{1},\bar \theta)
 \tilde{f}_2(y;\pi_{2},\bar \theta)  \tilde{f}_3(y;\pi_{3},\bar \theta)
  \nonumber\\
  ~ \nonumber\\
    && \! \! \!\!\!\!\! \! =(\hat{f}_1\cdot\hat{f}_2)\cdot\hat{f}_3
\eea
Moreover,   a mapping between the product
 $\hat f_1\cdot\hat f_2$ in deformed superspace
 and a star product of the corresponding functions in ordinary
 space can be
 established,
\be \hat{f}_1\cdot\hat{f}_2 = \int d^2\pi \exp(-\pi \hat \theta)
\widetilde{(f * g)}(y;\pi,\bar \theta) \ee
where the Moyal-Weyl star product is defined by
\bea
  f_1(y, \theta, \bar \theta) * f_2 (y, \theta, \bar
  \theta)&\equiv&
   f_1 (y, \theta, \bar \theta) \exp\left(-\frac {1}{2}C^{\alpha\beta}(y)
   \frac{\overleftarrow{\partial}}{\partial \theta^\alpha} \frac{
   \overrightarrow{\partial}}{\partial \theta^\beta} \right) f_2(y,
   \theta, \bar \theta) \nonumber\\
   &&= f_1 (y, \theta, \bar \theta) \left(1-\frac {1}{2}C^{\alpha\beta}(y)
   \frac{\overleftarrow{\partial}}{\partial \theta^\alpha} \frac{
   \overrightarrow{\partial}}{\partial \theta^\beta}-{\rm{det}}\, C(y)
   \frac{\overleftarrow{\partial}}{\partial \theta\theta} \frac{
   \overrightarrow{\partial}}{\partial \theta\theta} \right) f_2(y,
   \theta, \bar \theta)
   \label{producto}
\eea
with
\bea
    &&\frac{\overrightarrow{\partial}}{\partial \theta^\alpha}
    \theta^\beta \equiv \frac{{\partial}}
    {\partial \theta^\alpha}\theta^\beta = \delta^\beta_\alpha \nonumber\\
    &&\theta^\alpha \frac{\overleftarrow{\partial}}{\partial
    \theta^\beta}\equiv-\delta^\beta_\alpha \nonumber\\
    &&\frac{{\partial}}{\partial \theta\theta}\equiv \frac 1 4
    \epsilon^{\alpha\beta}\frac{{\partial}}{\partial \theta^\alpha}\frac{{\partial}}
    {\partial \theta^\beta}
\eea
Using the inverse Fourier transformation, we have
\bea    {\widetilde{{f}_1*{f}_2}}(y;\pi,\bar \theta)&=& \int
d^{2}\theta\,
   e^{\pi \theta} \left ({f}_1(y;\theta, \bar \theta)* {f}_2(y;\theta,\bar \theta)
   \right)\nonumber\\
   &=&\int d^{2}\theta\,e^{\pi \theta} \left (\int d^{2}
   \pi_1 e^{- \pi_1 \theta}\tilde{f_1} (y;\pi_1,\bar \theta)\right)*
   \left (\int d^{2} \pi_2 e^{- \pi_2 \theta}
   \tilde{f_2} (y;\pi_2,\bar \theta)\right)\nonumber\\
   &=&\int d^{2}\pi_1 d^{2} \pi_2 d^{2}\theta\,\tilde{f_1} (y;\pi_1,\bar \theta)
   \tilde{f_2}(y;\pi_2,\bar \theta)e^{\pi \theta} \left (e^{- \pi_1 \theta}*e^{-\pi_2 \theta}
   \right)\nonumber\\
   &=&\int d^{2}\pi_1 d^{2} \pi_2 d^{2}\theta\,\tilde{f_1} (y;\pi_1,\bar \theta)
   \tilde{f_2}(y;\pi_2,\bar \theta)e^{\pi \theta} e^{- (\pi_1 +\pi_2)\theta-
   \frac 1 2 \pi_1 C \pi_2} \nonumber\\
   &=&\frac 1 2\int d^{2}\pi_1 d^{2} \pi_2 \,\tilde{f_1} (y;\pi_1,\bar \theta)
   \tilde{f_2}(y;\pi_2,\bar \theta)
   \delta(\pi - \pi_1 -\pi_2)e^{-\frac 1 2 \pi_1 C \pi_2}
   \nonumber\\
   &=&\int d^{2}\pi'\,\tilde{f}_1 (y;\frac {\pi +\pi'}{2},\bar \theta) \tilde{f}_2
   (y;\frac {\pi -\pi'}{2},\bar \theta)e^{-\frac 1 8 (\pi +\pi')C(\pi
   -\pi')}
   \label{fourier2}
\eea
Thus, we see that ${\widetilde{{f}_1*{f}_2}}$ in (\ref{fourier2})
coincides with  ${\widetilde{{f}_1\cdot{f}_2}}$ in
(\ref{fourier1}), and then, $(f_1\cdot f_2)(\hat\theta)$ in
deformed superspace is mapped to $(f_1*f_2)(\theta)$ in the
ordinary superspace.

We  can then  formulate a field theory in the $C(y)$-deformed
superspace as defined above, by working in ordinary 4 dimensional
Euclidean superspace  but multiplying superfields with the
Moyal-Weyl product (\ref{producto}). Let us first discuss how the
algebra of the supercharges and covariant derivatives are modified
when a $y$-dependent $C$-deformation is introduced. Supercharges
and covariant derivatives in chiral coordinates take the form
\be
   Q_\alpha = \frac{\partial}{\partial \theta^\alpha} \; , \;\;\;\;\;
   \;\;\;\;\; \;\;\;\;\; \;\;\;\;\; {\bar Q}_{\dot \alpha} = -
   \frac{\partial}{\partial {\bar\theta}^{\dot\alpha}} + 2i
   \theta^\alpha \sigma^\mu_{\alpha\dot\alpha}
   \frac{\partial}{\partial y^\mu}, \ee \be {D}_{ \alpha} =
   \frac{\partial}{\partial {\theta}^{\alpha}} + 2i
   \sigma^\mu_{\alpha\dot\alpha}{\bar \theta}^{\dot\alpha}
   \frac{\partial}{\partial y^\mu}  \; , \;\;\; \;\; \;\;\;\;\;
   \;\;\;\;\; \;\;\;\;\; {\bar D}_{\dot \alpha} = -
   \frac{\partial}{\partial {\bar\theta}^{\dot \alpha}}
\ee
The covariant derivative algebra is not modified by (\ref{uno})
and the same happens for the  supercharge-covariant derivative
algebra. Concerning the supercharge algebra, it gets modified
according to
\bea
   \{Q_\alpha,Q_\beta\}_* &=& 0 \nonumber \\
   \{{\bar Q}_{\dot \alpha}, Q_\alpha \}_* &=&
   2i \sigma^\mu_{\alpha \dot \alpha} \frac{\partial}{\partial y^\mu} \nonumber \\
   \{ {\bar Q}_{\dot \beta}, {\bar Q}_{\dot \alpha}\}_* &=& -2
   \sigma^\mu_{\alpha \dot \alpha}\sigma^\nu_{\beta \dot \beta} ( \partial_\mu
   C^{\alpha\beta}\frac{\partial}{\partial y^\nu} + \partial_\nu C^{\alpha\beta}
   \frac{\partial}{\partial y^\mu } + 2 C^{\alpha \beta}
   \frac{\partial^2}{\partial y^\mu \partial y^\nu})
\eea
Only the chiral subalgebra generated by $Q_\alpha$ is still
preserved and this defines the chiral ${\cal N} = 1/2$
supersymmetry algebra. By a similar analysis, it can be seen that,
as it happens for constant $C^{\alpha\beta}$, the complete ${\cal
N}=1$ superconformal group is broken for the $y$-dependent
deformation to the subgroup generated by $\bar M_{\mu\nu},\;
D_{new}\equiv D- \frac 1 2 R,\; P_\mu, \; Q_\alpha, \; \bar
S^{\dot\alpha} $ which is known as ${\cal N} = 1/2$ superconformal
group.

\section{Scalar and Vector superfields}

A chiral superfield $\Phi$ satisfies the condition $\bar D_{\dot
\alpha}* \Phi = 0$. As usual, it can be written in the form
\be
   \Phi(y,\theta) = \phi(y) + \sqrt 2 \theta \psi(y) +
   \theta\theta F(y).
   \label{chiral}
\ee
The Moyal-Weyl product multiplication
 of two chiral superfields $\Phi_1 (y,\theta)$ and
$\Phi_2 (y,\theta)$  takes the form
\bea
   \Phi_1 (y,\theta) * \Phi_2 (y,\theta) &=& \Phi_1 (y,\theta)  \Phi_2 (y,\theta)
   -C^{\alpha\beta}(y)\psi_{1 \alpha}(y)\psi_{2
   \beta}(y) + \sqrt 2C^{\alpha\beta}(y) \theta_\beta(\psi_{1 \alpha}(y)F_2(y)-\psi_{2
   \alpha}(y)F_1(y)) \nonumber \\
   && - {\rm det} C (y) F_1(y)F_2(y).
\eea
It is easy to see  that the r.h.s. is a function   of $y$ and
$\theta$ solely, so that the product of two chiral superfields is
still a chiral superfield. This result could have been advanced
provided the covariant derivative $\bar D_{\dot\alpha}$ satisfies
the Leibniz rule. That this is the case can be seen by considering
 $\bar D_{\dot\alpha}$ acting on a product
of two superfields,
\bea
      \bar D_{\dot\alpha}(\Phi(y, \theta, \bar \theta)*\Psi(y,\theta, \bar
    \theta)) &=& \bar D_{\dot\alpha} \left [\Phi  \exp\left(-\frac {1}{2}C^{\alpha\beta}(y)
    \frac{\overleftarrow{\partial}}{\partial \theta^\alpha}
    \frac{\overrightarrow{\partial}}
    {\partial \theta^\beta} \right) \Psi\right ] \nonumber \\
    & = &\bar D_{\dot\alpha}(\Phi) \exp\left(-\frac {1}{2} C^{\alpha\beta}(y)
    \frac{\overleftarrow{\partial}}{\partial \theta^\alpha}
    \frac{\overrightarrow{\partial}}{\partial \theta^\beta} \right) \Psi
     + (-1)^{F[\Phi]}\Phi \exp\left(-\frac {1}{2}
    C^{\alpha\beta}(y) \frac{\overleftarrow{\partial}}{\partial \theta^\alpha}
    \frac{\overrightarrow{\partial}}{\partial \theta^\beta} \right) \bar D_{\dot\alpha}
    \Psi\nonumber \\
    & = &\bar D_{\dot\alpha}(\Phi) * \Psi + (-1)^{F[\Phi]}\Phi * \bar D_{\dot\alpha}\Psi
    \label{leib1}
\eea

The case of antichiral superfields is more involved. An antichiral
superfield is defined by $ D_{\alpha}* \bar\Phi = 0$. As usual,
$\bar \Phi$ only depends on $\bar \theta$ and the antichiral
coordinates $ {\bar y}^\mu = y^\mu - 2i \theta^\alpha
 \sigma^\mu_{\alpha \dot \alpha} {\bar \theta}^{\dot \alpha}$. Written in terms
of the chiral variable $y^\mu$, $\bar \Phi$  takes the form
\bea
   \bar \Phi(y-2i\theta\sigma\bar\theta,\bar\theta) &=& \bar\phi(y-2i\theta\sigma\bar\theta)
   + \sqrt 2\bar\theta\bar\psi(y-2i\theta\sigma\bar\theta)+\bar\theta\bar\theta\bar F(y
   -2i\theta\sigma\bar\theta) \nn\\
   &=&  \bar\phi(y) + \sqrt 2\bar\theta\bar\psi(y)-2i
   \theta\sigma^\mu\bar\theta\partial_\mu\bar\phi(y)+\bar\theta\bar\theta\left(\bar F(y)+ i\sqrt{2}\theta\sigma^\mu\partial_\mu\bar\psi(y)
   +\theta\theta\partial^\mu\partial_\mu\bar\phi\right).
\eea
The product of two antichiral superfields takes the form
\bea
     \bar \Phi_1(y-2i\theta\sigma\bar\theta,\bar\theta)*\bar
     \Phi_2(y-2i\theta\sigma\bar\theta,\bar\theta)\!\! &=& \!\! \bar \Phi_1(y-2i\theta\sigma\bar\theta,\bar\theta)\bar
     \Phi_2(y-2i\theta\sigma\bar\theta,\bar\theta) + 2\bar\theta\bar\theta C^{\mu\nu}(y) \partial_\mu \bar\phi_1(y)\partial_\nu
     \bar\phi_2(y)
\eea
Since the term $C^{\mu\nu}(y)\partial_\mu
\bar\phi_1(y)\partial_\nu \bar\phi_2(y)$ appears multiplied by
$\bar \theta\bar\theta$, its arguments can alternatively be taken
as the antichiral coordinate $\bar y^\mu$. It is then clear that
the product of two antichiral superfields is another antichiral
superfield. This result, however, turns out to be unexpected if
one notes that, due to the coordinate dependence of the
deformation $C^{\alpha\beta}$, the covariant derivative $D_\alpha$
violates the Leibniz rule. One can see this from  the product of
two generic superfields
\bea
    D_{\alpha}(\Phi(y, \theta, \bar \theta)*\Psi(y,\theta, \bar \theta))
    &=& D_{\alpha} \left [\Phi  \exp\left(-\frac {1}{2}C^{\alpha\beta}(y)
    \frac{\overleftarrow{\partial}}{\partial \theta^\alpha}
    \frac{\overrightarrow{\partial}}
    {\partial \theta^\beta} \right) \Psi\right ] \nonumber \\
    &=& D_{\alpha}(\Phi) \exp\left(-\frac {1}{2} C^{\alpha\beta}(y)
    \frac{\overleftarrow{\partial}}{\partial \theta^\alpha}
    \frac{\overrightarrow{\partial}}{\partial \theta^\beta} \right) \Psi +
    (-1)^{F[\Phi]}\Phi \exp\left(-\frac {1}{2}
    C^{\alpha\beta}(y) \frac{\overleftarrow{\partial}}{\partial \theta^\alpha}
    \frac{\overrightarrow{\partial}}{\partial \theta^\beta} \right) D_{\alpha}
    \Psi \nonumber \\
    &&+ \Phi D_{\alpha}\left [\exp\left(-\frac {1}{2}
    C^{\alpha\beta}(y) \frac{\overleftarrow{\partial}}{\partial \theta^\alpha}
    \frac{\overrightarrow{\partial}}{\partial \theta^\beta} \right)\right] \Psi\nonumber \\
    &=& D_{\alpha}(\Phi) * \Psi + (-1)^{F[\Phi]}\Phi * D_{\alpha}\Psi +
  \Phi D_{\alpha}(*) \Psi
    \label{leib2}
\eea
where we have introduced the notation $D_{\alpha}(*)$ to denote
\bea
    \Phi D_{\alpha}(*) \Psi = -2i(\sigma^\mu\bar\theta)_\alpha
    \left ( \frac {1}{2}\partial_\mu (C^{\beta\gamma})
   \Phi\frac{\overleftarrow{\partial}}{\partial \theta^\beta} \frac{
   \overrightarrow{\partial}}{\partial \theta^\gamma}\Psi+\partial_\mu
   ({\rm{det}}\, C) \Phi\frac{\overleftarrow{\partial}}
   {\partial \theta\theta} \frac{\overrightarrow{\partial}}
   {\partial \theta\theta}\Psi \right )
\eea

Let us now discuss vector superfields. We shall consider a
$U(N_c)$ gauge group with Lie algebra hermitian generators $T^a$
satisfying $[T^a,T^b] = i f^{abc} T^c$ and ${\rm tr} T^a T^b =
\frac{1}{2}\delta^{ab}$. In four dimensional ${\cal N}=1$
Euclidean superspace no reality condition analogous to the
Minkowski case can be imposed on  superfields. Hence, a vector
superfield $V$ containing the gauge field can be defined as the
one which changes under supergauge transformations according to
\be
   e^V \to e^{V'} = e^{-i\bar \Lambda} * e^V * e^{i \Lambda}
   \label{gauge0}
\ee
where $V= V^a T^a$ and $\Lambda= \Lambda^aT^a $ and $\bar\Lambda =
\bar \Lambda^a T^a$ are chiral and anti-chiral superfields taking
values in the Lie algebra of $U(N_c)$. In all the expressions
above exponentials are defined through their $*$-product
expansion,
\be
   e^{i\Omega} \equiv 1 + i \Omega + \frac{i^2}{2} \Omega * \Omega + \ldots.
\ee
Taking the standard expressions for the chiral and antichiral
superfields strength,
\bea
    W_\alpha &=& -\frac{1}{4} \bar D * \bar D * e^{-V} *
    D_\alpha * e^{V}\nonumber\\
    {\bar W}_{\dot \alpha} &=& \frac{1}{4} D * D * e^{V}* {\bar
    D}_{\dot\alpha}* e^{-V}
\eea
and using eqs.(\ref{leib1}),(\ref{leib2}) one verifies  that
 these superfields
transform under an infinitesimal supergauge
transformation according to
\bea
    \delta W_\alpha &=& i\left(W_\alpha * \Lambda - \Lambda *
    W_\alpha \right ) -\frac{i}{4} \bar D \bar D
    \left [ e^{-V}*\left(e^{V} D_\alpha(*)\Lambda - \bar \Lambda
    D(*) e^V\right )\right]\nonumber \\
    \delta \bar W_{\dot\alpha} &=& i\left(\bar W_{\dot\alpha} *
    \bar\Lambda - \bar\Lambda * \bar W_{\dot\alpha} \right )  +\frac{i}{2} D_\alpha(e^{V}* {\bar D}_{\dot\alpha} e^{-V})D^\alpha
    (*)\bar\Lambda - \frac{i}{2} \bar\Lambda D^\alpha(*)D_\alpha(e^{V}* {\bar D}_{\dot\alpha}
    e^{-V})
    \label{delta}
\eea
It is then clear that neither chiral nor antichiral superfield
strengths transform covariantly under a general supergauge
transformation unless the deformation parameter $C^{\alpha\beta}$
is constant. Gauge covariance cannot  be invoked  to transform an
arbitrary vector superfield to the Wess-Zumino gauge. One can
however still handle a $C=C(y)$ deformation if one starts  with a
vector
 superfield $V$ already satisfying the
Wess-Zumino condition. As in the case of  constant deformation, it
is convenient to identify the component fields in $V$ according to
\bea
    V(y,\theta.\bar\theta) &=& -\theta \sigma^\mu \bar \theta A_\mu(y)
    - i \bar\theta\bar\theta \theta^\alpha \left(\lambda_\alpha(y) + \frac 1 4
    \varepsilon_{\alpha\beta}C^{\beta\gamma}(y)
    \sigma^\mu_{\gamma\dot\gamma}\{{\bar \lambda}^{\dot
    \gamma}(y),A_\mu(y)\} \right)
    + i \theta\theta\bar\theta \bar \lambda(y)
    + \frac{1}{2} \theta\theta\bar\theta\bar\theta
    (D(y) - i \partial_\mu A^\mu(y))\nonumber\\ \label{wz}
\eea
In this gauge
\ba
   &&  V_*^2 \equiv V*V= -\frac{1}{2}\,
   \bar\theta\bar\theta\left[ \theta\theta A_\mu
    A^\mu + C^{\mu\nu}A_\mu A_\nu-\frac i 2
     \theta_\alpha C^{\alpha\beta}\sigma^\mu_{\beta\dot\beta}
    [A_\mu, \bar\lambda^{\dot\beta}] + \frac{1}{4}|C|^2\bar\lambda\bar\lambda\right]
    \nn\\
  &&  V_*^3 = 0
\ea
where $C^{\mu\nu}=C^{\alpha\beta}\varepsilon_{\beta\gamma}
(\sigma^{\mu\nu})_\alpha^{\,\,\gamma}$ is selfdual, and $|C|^2
\equiv C^{\mu\nu}C_{\mu\nu}=4\, {\rm det}C$.

Chiral and antichiral  superfield strengths, written in components
take the form
\bea
    W_\alpha &=& W_\alpha(C=0) +\varepsilon_{\alpha\beta}
    C^{\beta\gamma}\theta_\gamma\bar\lambda\bar\lambda \nonumber\\
    {\bar W}_{\dot \alpha} &=& {\bar W}_{\dot \alpha}(C=0) -
    \bar\theta\bar\theta\left [\frac {C^{\mu\nu}} {2}\{ F_{\mu\nu},
    \lambda_{\dot\alpha}\} + C^{\mu\nu} \{A_\nu,{\cal D}_\mu
    \bar\lambda_{\dot\alpha} -\frac i 4
    [A_\mu,\bar\lambda_{\dot\alpha}]\} +\frac {i} {16} |C|^2 \{\bar\lambda\bar\lambda, \bar\lambda_{\dot\alpha}
    \} + \partial_\mu C^{\mu\nu}\{\bar\lambda_{\dot\alpha},A_\nu\}\right ]
    \label{barw}
\eea
where
\bea
    F_{\mu \nu} &=& \partial_\mu A_\nu - \partial_\nu A_\mu +
    \frac i 2 [A_\mu, A_\nu] \nonumber \\
    {\cal D}_\mu \bar\lambda_{\dot\alpha} &=& \partial_\mu
    \bar\lambda_{\dot\alpha} + \frac i 2 [A_\mu, \bar\lambda_{\dot\alpha}]
\eea
One can still perform infinitesimal supergauge transformations
preserving the Wess-Zumino gauge (\ref{wz}) through
\ba
   \Lambda &=& -\varphi(y)\nn\\
   \bar\Lambda &=&-\varphi(y)
   +2i\theta\sigma^\mu\bar\theta\partial_\mu\varphi(y)
   -\theta\theta\bar\theta\bar\theta\partial^\mu\partial_\mu\varphi(y)
   -\frac i 2 \bar\theta\bar\theta C^{\mu\nu}\{\partial_\mu\varphi,A_\nu\}
   \label{gauge}
\ea
As in \cite{Seiberg}, a particular parametrization of the
coefficient of $\bar \theta \bar \theta\theta$ in $V$ is adopted
to ensure that the gauge transformation above acts on the
component fields in the standard way,
\bea
    \delta A_\mu &=& -2 \partial_\mu \varphi+i[\varphi,A_\mu]\nonumber\\
    \delta \lambda_\alpha &=& i[\varphi,\lambda_\alpha]\nonumber\\
    \delta{\bar \lambda}_{\dot \alpha} &=& i[\varphi,{\bar \lambda}_{\dot \alpha}]\nonumber\\
    \delta D &=& i[\varphi,D]
    \label{gauge1}
\eea

For the case of the supergauge transformation (\ref{gauge}), the
transformation of the superfield strengths reduces to
\bea
    \delta W_\alpha &=& i\left(W_\alpha * \Lambda - \Lambda *
    W_\alpha \right ) \nonumber \\
    \delta \bar W_{\dot\alpha} &=& i\left(\bar W_{\dot\alpha}  *
    \bar\Lambda - \bar\Lambda * \bar W_{\dot\alpha} \right ) + 2 \bar\theta\bar\theta
    \partial_\mu
    C^{\mu\nu}\{\bar\lambda_{\dot\alpha},\partial_\nu\varphi\}
    \label{ww2}
\eea
Finally,  gauge covariance under the set of transformations
(\ref{gauge}) is achieved provided the condition
\be
   \partial_\mu C^{\mu\nu}=0
   \label{condition}
\ee
holds.

As discussed  for the case of constant deformations in
\cite{deBoer}-\cite{Ber1}, $C^{\mu\nu}$  is related to the
 graviphoton field-strength background through the
 formula $(\alpha')^2
F^{\mu\nu} =
 C^{\mu\nu}$ and   $F^{\mu\nu}$ is taken  as selfdual in order to
 avoid back reaction of the metric. It is then natural to
 interpret
 condition (\ref{condition}), that in our approach follows
 from gauge covariance arguments,    as the graviphoton equation
 of motion   provided the relation between
 the coordinate dependent deformation and a self-dual  graviphoton field
remains valid.
 Concerning supersymmetry, one easily checks, following the
analysis in \cite{Ber1},  that a coordinate dependent self-dual
graviphoton background does not affect the 4 chiral supercharges
$Q_\alpha$. In principle, the antichiral supercharges will be
broken for a nonconstant background (but this should be
investigated more thoroughly). As we shall see below from the
Lagrangian  written in components,  condition (\ref{condition})
guarantees the gauge invariance of the theory.

  ~

Let us end this section with a comment on supersymmetry
transformations. Infinitesimal SUSY transformations are generated
by the operator $Q_\alpha=\partial / \partial_\alpha$ acting on
superfields. As it is well known, this operation takes the vector
superfield away from the W-Z gauge. For consistency, we have to be
able to restore the W-Z gauge in $V$ after the SUSY
transformation, while maintaining supergauge invariance of ${\rm
tr}\,W^2$ and ${\rm tr}\,\bar W^2$. The supergauge transformation
restoring the W-Z gauge is in the present case generated by
\be
   \Lambda=0 \, , \;\;\;\;\;\;\;\;\;\;\;\;\; \bar\Lambda= i \xi
   \sigma^\mu \bar \theta A_\mu - \bar\theta\bar\theta \left(
   \xi \lambda - \frac{1}{2}\xi \sigma_\mu\bar \lambda C^{\mu\nu}
   A_\nu\right)
   \label{gauge2}
\ee
It can be seen from (\ref{delta}) that both $W$ and $\bar W$
transform covariantly under the supergauge transformation
(\ref{gauge2}). Such composition of SUSY and gauge tranformations
gives
\bea
   && \delta A_\mu = - i \bar \lambda {\bar \sigma}_\mu \xi  \;\;\;
    \rightarrow  \;\;\;
    \delta F_{\mu\nu} = i\xi(\sigma_\nu {\cal D}_\mu
    - \sigma_\mu {\cal D}_\nu )\bar \lambda\nonumber\\
   && \delta \lambda = i D \xi+
    \sigma^{\mu\nu}\xi\left( F_{\mu\nu}
    +\frac i 2\, C_{\mu\nu}\bar \lambda \bar \lambda
    \right)\nonumber \\
    &&\delta \bar \lambda= 0 \nonumber \\
   && \delta D =-\xi\sigma^\mu {\cal D}_\mu \bar \lambda
    \label{bogo2}
\eea

\section{Euclidean ${\cal N}=1/2$ SQCD}

From the results above, we see that once condition
(\ref{condition}) is imposed, it should be possible to
consistently construct a Lagrangian
 invariant both under generic supersymmetric  and
particular supergauge transformations which correspond to the
standard transformations of the component fields. The super
Yang-Mills Lagrangian in $C(y)$-deformed superspace takes the form
\bea
    {\cal L}^{SYM}=\frac{1}{8g^2}\left({\rm tr}\!\!\int\! d^2\theta\,
    W^\alpha*W_\alpha + {\rm tr}\!\!\int\! d^2\bar\theta\, \bar
    W_{\dot\alpha}*\bar W^{\dot\alpha}\right)
\eea
Using the expressions (\ref{barw}) for the superfield strengths,
the F-terms  are
\bea
  {\mathrm{tr}}\; W^\alpha * W_\alpha\mid_{\theta \theta} &=& {\mathrm{tr}}\;
  W^\alpha W_\alpha (C=0)\mid_{\theta   \theta} - i C^{\mu\nu}
  {\mathrm{tr}}F_{\mu\nu}\bar\lambda\bar\lambda +
  \frac {|C|^2} {4} {\mathrm{tr}} (\bar\lambda\bar\lambda )^2
  \nonumber
  \\{\mathrm{tr}}\; \bar W_{\dot\alpha} * \bar W^{\dot\alpha}\mid_{\bar\theta \bar\theta}
  &=&
  {\mathrm{tr}}\; \bar W_{\dot\alpha} \bar W^{\dot\alpha}(C=0)\mid_{\bar\theta \bar\theta} -
  i C^{\mu\nu}{\mathrm{tr}}F_{\mu\nu}\bar\lambda\bar\lambda +
  \frac {|C|^2} {4} {\mathrm{tr}} (\bar\lambda\bar\lambda )^2 -2i {\mathrm{tr}}\partial_\mu(C^{\mu\nu}A_\nu\bar\lambda\bar\lambda)
  \label{ww}
\eea

Then, disregarding the surface term, the ${\cal N}=1/2$ super
Yang-Mills Lagrangians in term of the component fields is
\be
    {\cal L}_C^{SYM} = {\cal L}_{C=0}^{SYM}-
  \frac {i}{4g^2} C^{\mu\nu}(y){\mathrm{tr}}F_{\mu\nu}\bar\lambda\bar\lambda +
  \frac {1} {16g^2}|C(y)|^2 {\mathrm{tr}} (\bar\lambda\bar\lambda )^2
  \label{langa}
\ee
which coincides with the usual expresion for ${\cal N}=1/2$
deformed Super Yang-Mills theory for constant $C^{\alpha \beta}$
\cite{Seiberg}.

Let us now  add matter fields in order to consider a
supersymmetric version of QCD with $U(N_c)$ gauge group and $N_f$
flavors  defined in the deformed superspace. The matter fields
 are pairs
 of chiral superfields $\{\Phi, \Psi\}$ transforming as
$\{{\bf N_c},{\bf \bar N_c}\}$ multiplets of the colour group. The
super QCD (SQCD) Lagrangian is defined as
\be
   {\cal L}^{SQCD}={\cal L}^{SYM}_C + {\cal L}^{matter}_C
   \label{lag}
\ee
where we have redefined the  SYM Lagrangian in order to
incorporate a $\theta$-term,
\be
   {\cal L}^{SYM}_C= \frac {-i \tau}{32\pi}\int\! d^2\theta\,
    {\rm tr}\, W^\alpha*W_\alpha + \frac {i\bar \tau}{32\pi}\int\! d^2\bar\theta\,
    {\rm tr}\,\bar W_{\dot\alpha}*\bar W^{\dot\alpha}
\ee
with
 \be
\tau= \frac{\theta}{2\pi} + \frac{4\pi i}{g^2} \ee
 and $\bar\tau$ its complex conjugate. Concerning the matter Lagrangian,
\bea
  {\cal L}^{matter}_{ C} &=& \!\!\int\! d^4\theta\,\left (\bar\Phi * e^{V} *\Phi +
  \Psi * e^{-V} *\bar\Psi \right)  +\!\! \int\! d^2\theta\, m\,
  \Psi*\Phi + \!\!\int\! d^2\bar\theta\,\bar m\,
  \bar\Phi*\bar\Psi
  \label{lagmat}
\eea

The matter Lagrangian (\ref{lagmat}) is invariant under local
$U(N_c)$ supergauge transformations,
\bea
    (\Phi,\bar\Phi)&\rightarrow&(e^{-i\Lambda}*\Phi,\bar\Phi* e^{i\bar\Lambda}),\nonumber\\
    (\Psi,\bar\Psi)&\rightarrow&(\Psi* e^{i\Lambda},e^{-i\bar\Lambda}*\bar\Psi ),\nonumber\\
    e^{V}&\rightarrow& e^{-i\bar\Lambda}*e^{V}*e^{i\Lambda},
\eea

In order to have the ordinary gauge transformation laws for the
component fields under the supergauge transformation generated by
(\ref{gauge}), we parameterize the matter superfields as in
\cite{Araki},
\bea
   &&\Phi(y,\theta) = \phi(y) + \sqrt 2 \theta \psi(y) +
   \theta\theta F_{\phi}(y)\nonumber\\
   &&\bar \Phi(\bar y, \bar\theta) = \bar\phi(\bar y) +
    \sqrt 2\bar\theta\bar\psi(\bar y)
   +\bar\theta\bar\theta\left(\bar F_{\bar\phi}
   +iC^{\mu\nu}\partial_\mu\left(\bar \phi A_\nu\right)
   -\frac 1 4 C^{\mu\nu}\bar\phi A_\mu A_\nu \right)(\bar y)\nonumber\\
   &&\Psi(y,\theta) = \eta(y) + \sqrt 2 \theta \chi(y) +
   \theta\theta F_{\eta}(y)\nonumber\\
   &&\bar \Psi(\bar y, \bar\theta) = \bar\eta(\bar y) +
   \sqrt 2\bar\theta\bar\chi(\bar y)
   +\bar\theta\bar\theta\left(\bar F_{\bar\eta}
   +iC^{\mu\nu}\partial_\mu\left( A_\nu\bar \eta\right)
   -\frac 1 4 C^{\mu\nu} A_\mu A_\nu \bar\eta\right)(\bar y)\nonumber\\
\eea
Written in components, infinitesimal transformations read
\bea
    \delta\phi = i \varphi \phi \;\;\; && \;\;\; \delta\bar \phi
    = -i \bar \phi \varphi \nonumber\\
    \delta\psi = i \varphi \psi \;\;\; && \;\;\;  \delta\bar\psi = -i
    \bar\psi  \varphi \nonumber\\
    \delta F_{\phi} = i \varphi F_{\phi} \;\;\;  && \;\;\; \delta \bar F_{\bar\phi} =  -i \bar
    F_{\bar\phi} \varphi\nonumber \\
    \nonumber \\
    \delta\bar\eta = i \varphi \bar\eta \;\;\; && \;\;\; \delta\eta
    = -i \eta \varphi \nonumber\\
    \delta\bar\chi = i \varphi \bar\chi \;\;\; && \;\;\;  \delta\chi = -i
    \chi  \varphi \nonumber\\
    \delta \bar F_{\bar\eta} = i \varphi \bar F_{\bar\eta} \;\;\;  && \;\;\;
    \delta  F_{\eta} =  -i    F_{\eta} \varphi
    \label{transfor}
\eea
The different terms in the matter Lagrangian written in components
fields read
\bea
    m\,\Psi*\Phi\mid_{\theta\theta}
    &=&m\eta F_\phi + m F_\eta \phi -m\chi\psi \\
    \bar m\,\bar\Phi*\bar\Psi\mid_{\bar\theta
    \bar\theta}&=&\bar m\bar\phi\bar F_{\bar\eta}
    +\bar m \bar F_{\bar\phi}\bar\eta-\bar
    m\bar\psi\bar\chi+\frac i 2 \bar m C^{\mu\nu}\bar\phi F_{\mu\nu}\bar \eta
    +i\bar mC^{\mu\nu}\partial_\mu\left( \bar \phi A_\nu\bar \eta\right)
    +2\bar m C^{\mu\nu}\partial_\mu\bar\phi\partial_\nu\bar\eta
\eea
\bea
     \bar\Phi * e^{V} *\Phi\mid_{\theta\theta\bar\theta \bar\theta}
     + \Psi * e^{-V} *\bar\Psi\mid_{\theta\theta\bar\theta \bar\theta}&=&
      \vphantom{\frac {\sqrt2}{2}}\bar\Phi  e^{V} \Phi (C=0)\mid_{\theta\theta\bar\theta \bar\theta}
     + \Psi  e^{-V} \bar\Psi(C=0)\mid_{\theta\theta\bar\theta \bar\theta}\nonumber\\
    && + \frac i 2 C^{\mu\nu}\bar \phi F_{\mu\nu} F_{\phi} - \frac {1} {16} \,|C|^2
    \bar \phi\bar \lambda \bar \lambda F_{\phi }   -\frac {\sqrt2}{2}\,
    \overline{{D}_\mu  \phi} (\sigma^\mu \bar \lambda)_\alpha C^{\alpha \beta}
    \psi_{\beta}\nonumber\\
    && + \frac i 2 C^{\mu\nu}F_{\eta} F_{\mu\nu}\bar \eta  - \frac {1} {16} \,|C|^2
    F_{\eta}\bar \lambda \bar \lambda \bar \eta   -\frac {\sqrt2}{2}\,
    \chi_\alpha C^{\alpha \beta}(\sigma^\mu \bar \lambda)_{\beta}{{D}_\mu  \bar\eta}
\eea
where
\bea
    \bar\Phi  e^{V} \Phi (C=0)\mid_{\theta\theta\bar\theta
    \bar\theta}&=&\bar F_{\bar\phi} F_\phi - i{\bar \psi}\bar\sigma^\mu
    {D}_\mu \psi -\overline{{D}_\mu \phi} {D}^\mu \phi +\frac 1 2 \bar \phi D\phi
    + \frac {i}{\sqrt2} (\bar \phi \lambda \psi - \bar \psi \bar \lambda \phi)
    \\
    \Psi  e^{-V} \bar\Psi(C=0)\mid_{\theta\theta\bar\theta
    \bar\theta}&=& F_\eta \bar F_{\bar\eta} - i{\chi}\sigma^\mu
     {D}_\mu \bar\chi   -\bar D_\mu \eta {D}^\mu \bar\eta -\frac 1 2 \eta D\bar \eta
    + \frac {i}{\sqrt2} (\eta \bar\lambda \bar\chi - \chi  \lambda \bar\eta)
\eea
and
\bea
    &&{{D}_\mu\phi} = \partial_\mu \phi  + \frac i 2 A_\mu \phi\;\;\; \;\;\;
     \overline{{D}_\mu\phi} = \partial_\mu \bar\phi
     - \frac i 2 \bar\phi A_\mu  \;\;\; \;\;\;{D_\mu  \psi}= \partial_\mu \psi  +
    \frac i 2 A_\mu\psi\nonumber\\
    &&{D_\mu  \bar\eta}= \partial_\mu \bar\eta  +
    \frac i 2 A_\mu\bar\eta\;\;\; \;\;\;{\bar D_\mu  \eta}= \partial_\mu \eta
    - \frac i 2 \eta A_\mu \;\;\;\;\;\;{D_\mu  \bar\chi}= \partial_\mu \bar\chi  +
    \frac i 2 A_\mu\bar\chi\nonumber\\
\eea
Putting all this together, the Euclidean ${\cal N}=1/2$ SQCD
Lagrangian, in components takes the form
\bea {\cal L}^{SQCD} = {\cal L}^{SQCD}_{C = 0} + \sum_{i=1}^6
{\cal L}_i \label{lagarca}
 \eea
 \bea
 {\cal L}_1 &=& -
  \frac {i}{8g^2} C^{\mu\nu}(y){\mathrm{tr}}F_{\mu\nu}\bar\lambda\bar\lambda +
  \frac {1} {32g^2}|C(y)|^2 {\mathrm{tr}} (\bar\lambda\bar\lambda )^2\nonumber\\
{\cal L}_2   & = &  \frac i 2 C^{\mu\nu}(y)\bar \phi F_{\mu\nu}
F_{\phi} -
   \frac {1} {16} \,|C(y)|^2
  \bar \phi\bar \lambda \bar \lambda F_{\phi}  \nonumber\\
   {\cal L}_3  & = &  -\frac {\sqrt2}{2}\,
    \overline{{D}_\mu  \phi} (\sigma^\mu \bar \lambda)_\alpha C^{\alpha \beta}(y)
    \psi_{\beta}  \nonumber\\
      {\cal L}_4 &=&  \frac i 2 C^{\mu\nu}(y) F_{\eta} F_{\mu\nu}\bar \eta
    -\frac {1} {16} \,|C(y)|^2
    F_{\eta}\bar \lambda \bar \lambda \bar \eta \nonumber\\
     {\cal L}_5 &=&  -\frac {\sqrt2}{2}\,
   C^{\alpha \beta}(y) \chi_\alpha (\sigma^\mu \bar \lambda)_{\beta}{{D}_\mu
    \bar\eta} \nonumber\\
    {\cal L}_6 &= & \frac{i\bar m}{2}  C^{\mu\nu}(y) \bar\phi F_{\mu\nu}\bar \eta
  \label{lagc}
\eea
In obtaining (\ref{lagc})  we have used eq.(\ref{condition}) and
discarded surface terms. Auxiliary fields can be eliminated using
their equations of motion,
\bea
F_\phi &=& - \bar m \bar \eta \nonumber\\
\bar F_{\bar \phi} &=& - m \eta - \frac{i}{2} C^{\mu\nu} \bar \phi
F_{\mu\nu}
+ \frac{1}{16}|C|^2 \bar \phi \bar \lambda \bar \lambda  \nonumber\\
F_\eta &=& -\bar m \bar \phi \nonumber\\
\bar F_{\bar \eta} &=& - m \phi - \frac{i}{2}
C^{\mu\nu}F_{\mu\nu}\bar \eta + \frac{1}{16} |C|^2 \bar \lambda
\bar \lambda \bar \eta \label{auxi} \eea

The ${\cal N}=1/2$ supersymmetric variations of the matter
component fields under which this Lagrangian is invariant are
\cite{Araki}
\bea
   && \delta \phi = \sqrt 2 \xi \psi,  \;\;\;\;\;\;\delta \bar \phi =0,\;\;\;
   \;\;\;\delta \psi = \sqrt 2 \xi F_\phi,\;\;\;\;\;\;\delta \bar \psi_{\dot\alpha}
   = -i\sqrt 2 \overline{D_\mu\phi}(\xi\sigma^\mu)_{\dot\alpha},\nonumber
   \\&&\delta F_\phi = 0, \;\;\;\;\;\;\;\;\;\;\;\; \delta \bar F_{\bar\phi} = -i \sqrt 2\,
   \overline{D_\mu \psi} {\bar\sigma}^\mu \xi -i\bar \phi \xi \lambda +
   C^{\mu\nu}\bar D_\mu\left( \bar \phi \xi \sigma_\nu\bar
   \lambda\right),\nonumber
   \\&& \delta \eta = \sqrt 2 \xi \chi,  \;\;\;\;\;\;\delta \bar \eta =0,\;\;\;
   \;\;\;\delta \chi = \sqrt 2 \xi F_\eta,\;\;\;\;\;\;\delta \bar \chi_{\dot\alpha}
   = -i\sqrt 2 D_\mu\bar\eta(\xi\sigma^\mu)_{\dot\alpha},\nonumber
   \\&&\delta F_\eta = 0, \;\;\;\;\;\;\;\;\;\;\;\; \delta \bar F_{\bar\eta} = -i \sqrt 2\,
   D_\mu \bar\chi {\bar\sigma}^\mu \xi -i \xi \lambda\bar \eta +
   C^{\mu\nu} D_\mu\left( \xi \sigma_\nu\bar\lambda\bar \eta
   \right).
   \label{bogo22}
\eea
It was pointed out in \cite{I} that the variation of the
$C$-deformed Super Yang-Mills Lagrangian (with constant $C_{\alpha
\beta}$) can be written as a $Q$ commutator. Then, if
supersymmetry is not spontaneously broken,
 the partition function and, in general, correlation functions
of $Q$ invariant operators will not depend on $C$. The extension
  to the
case in which massless matter fields are present, was considered
in \cite{Giombi}, also for the constant $C$ case.  One can see
that such a formal analysis can be done in the case of ${\cal
N}=1/2$ SQCD Lagrangian (\ref{lagc}) with $C = C(y)$. Indeed,
after some straightforward calculations one finds
\bea
\frac{\delta {\cal L}_1}{\delta C^{\mu\nu}} &=&    \frac
  {i}{16g^2}{\rm tr}\{Q^\alpha,(\sigma_{\mu\nu})_{\alpha\beta}\lambda^\beta
  \bar\lambda\bar\lambda\}
\nonumber\\
 \frac{\delta {\cal L}_2}{\delta C^{\mu\nu}}  &=& \frac
    {i}{4}\{Q^\alpha,\bar\phi(\sigma_{\mu\nu})_{\alpha}^{\,\beta}\lambda_\beta
    F_\phi\ +i\frac{\sqrt 2}{8}  C_{\mu\nu}\bar\phi
    \bar\lambda\bar\lambda\psi_\alpha\}\nonumber\\
    \frac{\delta {\cal L}_3}{\delta C^{\mu\nu}} &=&
    \frac {i}{4}\{Q^\alpha,\bar\psi\bar\lambda(\sigma_{\mu\nu})_{\alpha}^{\,\beta}
    \psi_\beta\}\nonumber\\
   \frac{\delta {\cal L}_4}{\delta C^{\mu\nu}} &=&    \frac
    {i}{4}\{Q^\alpha,F_\eta(\sigma_{\mu\nu})_{\alpha}^{\,\beta}\lambda_\beta
    \bar\eta +i \frac {\sqrt 2}{8} C_{\mu\nu}\chi_\alpha
    \bar\lambda\bar\lambda\bar\eta\}\nonumber\\
    \frac{\delta {\cal L}_5}{\delta C^{\mu\nu}} &=&
    -\frac {i}{4}\{Q^\alpha,(\sigma_{\mu\nu})_{\alpha}^{\,\beta}\chi_\beta\bar
    \lambda\bar\chi\}\nonumber\\
   \frac{\delta {\cal L}_6}{\delta C^{\mu\nu}} &=&
   \frac{\bar m}{2} \{Q^\alpha,\bar\phi
    (\sigma_{\mu\nu})_{\alpha}^{\,\beta}\lambda_\beta -
    \frac{1}{4}  C_{\mu\nu}\bar\phi A_\rho (\sigma^\rho\bar\lambda)_\alpha\bar\eta
    \}
    \label{todo}
\eea
It should be stressed that  SUSY transformations (\ref{bogo22})
were used in order to write the different $C$-dependent terms in
the Lagrangian as  $Q$ commutators. To confirm that the connection
still holds at the quantum level one should analyze  whether no
anomalous contributions modify the classical identities. To this
end, one should proceed to a calculation similar to that presented
at the end of this section in the analysis of Konishi anomaly and
gluino condensation. We leave the details of the complete analysis
confirming these identities for a forthcoming work.

 At the formal
level, if we  assume that supersymmetry is not spontaneously
broken and hence $Q|0\rangle = 0$,
 we can then write
 \be
 \frac{1}{Z}\frac {\delta Z}{\delta C^{\mu\nu}(y)} = 0
\ee
with
\be
Z = \int D fields \exp\left(-\int d^4x {\cal L}^{SQCD}\right)
\label{ri}
\ee

\subsection*{Konishi anomaly and gluino condensation}
Given the Lagrangian (\ref{lagarca}), it is  easy to verify that
the anomalous commutator
 leading to the gluino
condensation in the ${\cal N}=1/2$  supersymmetric theory
with coordinate dependent deformation gives
the same answer as in the undeformed case. That is, we shall show
that the following relation holds,

\be \frac{1}{2\sqrt 2} \{Q^\alpha, \chi_\alpha(y) \bar \eta(y)  \}
= - \bar m \bar \phi \bar \eta(y) + \frac{g^2}{32\pi^2} \bar
\lambda\bar \lambda(y) \label{gary} \ee
where the last term, corresponding to the Konishi anomaly
\cite{konishi}-\cite{konishi2}, results from a consistent
regularization of the ill-defined product in the commutator
\cite{sun}. Consider for instance the point spliting method where
one defines
\bea \left.\frac{1}{2\sqrt 2} \{Q^\alpha, \chi_\alpha(y) \bar
\eta(y)  \}\right\vert_{reg} &&\equiv \lim_{\epsilon \to 0}
\frac{1}{2\sqrt 2} \{Q^\alpha, \chi_\alpha(y+\epsilon) \exp(-i
\epsilon^\mu A_\mu)
 \bar \eta(y-\epsilon)  \}
\nonumber\\
&& = \left.
 \{Q^\alpha,
\chi_\alpha(y) \bar \eta(y)  \}\right\vert_{naive} + \label{ufa}
\lim_{\epsilon \to 0} \epsilon^\mu \chi_\alpha (y +
\epsilon)\sigma_\mu^{\alpha \dot\alpha} \bar\lambda_{\dot
\alpha}(y) \bar \eta(y - \epsilon) \eea
When inserted in a correlation function containing a product of
local operators, the second term in the r.h.s. of (\ref{ufa})
gives a finite contribution in the $\epsilon \to 0$ limit. This is
a due to a contribution from a linear ultraviolet divergent term
that results from the integration of a loop containing propagators
that arise from contractions with the Yukawa interaction term
 $\eta \bar \lambda \bar\chi$ present in ${\cal L}_{C=0}^{SQCD}$. The final answer
 coincides with eq.(\ref{gary}).  One can also  check that
 the new $C(y)$-dependent
 vertices in
 Lagrangian (\ref{lagc})-(\ref{auxi}) do not give rise to new  finite
 contributions
so that eq.(\ref{gary})
 holds for $C(y) \ne 0$ as it does in undeformed superspace.

\section{Summary and discussion}

Our work was motivated by the observation in \cite{shifman}
relating the spectral degeneracy in conventional ${\cal N}=1$ SUSY
gluodynamics with a $C$ deformation of the anticommuting
superspace coordinates,  suggesting that  ${\cal N}=1/2$
supersymmetry might be defined  for a coordinate dependent $C$
parameter.

In contrast with the case of ordinary noncommutative geometry,
where implementation of an associative $*$-product becomes rather
complicated for space-time dependent $\theta_{\mu\nu}(x)$ \cite{1}
(see also \cite{2} and references therein),  the case in which
$C_{\alpha\beta}$ depends on the chiral variable $y$
 can be rather simply handled and a
Moyal-Weyl star product can be defined (according to
eq.~\ref{producto}) so that associativity and other basic
properties remain valid (see also \cite{gracia}). One can then see that the subalgebra
generated by $Q_\alpha$ is still preserved and hence, as in the
constant $C$ case a chiral ${\cal N} = 1/2$ supersymmetry can be
defined with the superconformal group broken to the so-called
${\cal N} = 1/2$ superconformal group. Multiplication of chiral
superfields proceed as in the constant $C$ case while that of
antichiral ones is more involved because the Leibnitz rule for the
derivative of a product ceases to be valid. Concerning vector
superfields, a remarkable condition arises when studying the
covariance of superfield strengths, namely $\partial_\mu
C^{\mu\nu} =0$, which is consistent with the requirement of
selfduality of the graviphoton field background  present in the
supergravity model at the origin of (constant) C-deformations.

With all these ingredients, a ${\cal N}=1/2$ SQCD  Lagrangian was
constructed and, from its expression in components, the effects of
the deformation were discussed. In particular, studying the
Konishi anomaly we confirmed that, as suggested in \cite{shifman},
the anomalous commutator contribution leading to gluino
condensation has the same form as in the ordinary case.

Various interesting issues related to our work can be envisaged.
In particular one should analyze whether anomalous terms arise
when computing at the quantum level commutators like those in
eq.(\ref{todo}) as it happens in (\ref{ufa}) \cite{konishi}.
Another line to pursue concerns the corrections introduced by the
coordinate dependent deformation on BPS equations,  as it was
already discussed for constant $C$ in \cite{I2}-\cite{Ald}. We
hope to come back to  these issues and those related to the
connection with string theory dynamics in non-trivial graviphoton
background elsewhere.

\vspace{1.5 cm}

\noindent\underline{Acknowledgements}: We wish to thank J.M.
Maldacena and C.~N\'u\~nez for helpful comments. L.A. is supported
by CONICET. This work is partially supported by UNLP, CICBA, and
CONICET, ANPCYT (PICT grant 03-05179).

\end{document}